\documentclass[prl,showpacs,floatfix,superscriptaddress,amsmath,amssymb,twocolumn]{revtex4-1}
\usepackage{graphics}
\usepackage{graphicx}
\usepackage{amssymb}
\usepackage{amsmath}
\usepackage{epstopdf}
\usepackage{color}
\usepackage[colorlinks,bookmarks=false,citecolor=blue,linkcolor=red,urlcolor=blue]{hyperref}
\usepackage{times}
\usepackage{pdfpages}

\begin{document}
\newcommand\be{\begin{equation}}
\newcommand\ee{\end{equation}}
\newcommand\etal{{\it et al.}}

\newcommand{\ket}[1]{|{#1}\rangle}
\newcommand{\bra}[1]{\langle{#1}|}
\newcommand{\braket}[1]{\langle{#1}\rangle}
\newcommand{\ad}{a^\dagger}
\newcommand{\e}{\ensuremath{\mathrm{e}}}
\newcommand{\norm}[1]{\ensuremath{| #1 |}}
\newcommand{\aver}[1]{\ensuremath{\big<#1 \big>}}

\title{Effect of Rare Fluctuations on the Thermalization of Isolated Quantum Systems}
\author{Giulio Biroli}
\affiliation{Institut de Physique Th{\'e}orique, CEA/DSM/IPhT-CNRS/URA 2306 CEA-Saclay,
F-91191 Gif-sur-Yvette, France}
\author{Corinna~Kollath}
\affiliation{Centre de Physique Th\'eorique, CNRS, \'Ecole Polytechnique, 91128 Palaiseau Cedex, France}
\author{Andreas M. L\"auchli}
\affiliation{Max Planck Institut f\"ur Physik komplexer Systeme, D-01187 Dresden, Germany}

\date{\today}
\begin{abstract}
We consider the question of thermalization for isolated quantum systems after a sudden parameter change, a so-called quantum quench. In particular we investigate the pre-requisites for thermalization focusing on the statistical properties of the time-averaged density matrix and of the expectation values of observables in the final eigenstates. We find that eigenstates, which are rare compared to the typical
ones sampled by the micro-canonical distribution, are responsible for the absence of thermalization of 
some infinite integrable models and play an important role for some non-integrable systems of finite size, such as the Bose-Hubbard model. 
We stress the importance of finite size effects for the thermalization of isolated quantum systems and discuss two 
alternative scenarios for thermalization, as well as ways to prune down the correct one.
\end{abstract}

\maketitle
The microscopic description of many particle systems is very involved. In many situations, in particular at equilibrium, one can rely on statistical ensembles that provide a framework to compute time-averaged observables and obtain general results like fluctuation-dissipation relations. The use of statistical ensembles relies on the hypothesis that on long timescales physical systems thermalize. In classical statistical physics a very good understanding of thermalization was reached in the last century \cite{CastiglioneVulpiani2008}: under certain chaoticity conditions, an isolated system thermalizes at long times within the micro-canonical ensemble. Furthermore, a large single portion of a (much larger) isolated system thermalizes within the grand-canonical ensemble.
Instead for quantum systems, 
it is fair to state that the comprehension of thermalization and its 
pre-requisites are still open problems \cite{Reimann2008, GoldsteinZanghi2009}, except for important results obtained in the semi-classical limit \cite{Srednicki1994, Nonnenmacher2008} or for the coupling to a thermal bath \cite{Weissbook2008} \cite{footnote1}. And this is the case despite 
a lot of effort especially in the mathematical physics literature starting from the Quantum Ergodic Theorem of von~Neumann \cite{VonNeumann1929} (see  \cite{GoldsteinZanghi2009} for a very recent account and new results).\\
The interest in these fundamental questions revived recently due to their direct relevance for experiments in ultracold atomic gases \cite{BlochZwerger2008}. The almost perfect decoupling of these gases from their environment enables the investigation of the quantum dynamics of isolated systems. 
In a fascinating experiment by Kinoshita et al.~\cite{KinoshitaWeiss2006} it was observed that two counter-oscillating clouds of bosonic atoms confined in a one-dimensional harmonic trapping potential relax to a state different from the thermal one. 
Up to now the absence of thermalization \cite{footnote2} has been mainly attributed to the presence of infinitely many conserved quantities, i.e.~to the integrability of the system (see \cite{CalabreseCardy2007} and references therein). For non-integrable isolated models the presence of thermalization after a global quench, i.e. a sudden global parameter change, is still debated \cite{KollathAltman2007, ManmanaMuramatsu2007,CramerEisert2008,MoeckelKehrein2008,Roux2009, EcksteinWerner2009,Rigol2009}. 
The origin of thermalization (and its absence) after a global quench was proposed to stem from statistical properties of the time averaged density matrix and the so-called 'eigenstate-thermalization hypothesis' (ETH)~\cite{Deutsch1991,Srednicki1994,RigolOlshanii2008,Rigol2009}. ETH, roughly speaking, says that 
all eigenstates with the same intensive energies are thermal, meaning that expectation values of all local 
observables within the eigenstate coincide with the ones in the corresponding Gibbs ensemble (see later for a precise definition).   \\
The aim of our work is to understand to what extent ETH is a necessary and sufficient condition for thermalization. 
ETH can be interpreted in two different ways: a weak one which we show to be verified even for integrable models and which states that the fraction of the non-thermal states vanishes in the thermodynamic limit, and a strong one which states 
that non-thermal states completely disappear in the thermodynamic limit. The former interpretation does not imply thermalization. The reason is the possible existence of rare non-thermal states that can have a high overlap with the initial condition for the dynamics. We shall show that this is the origin of non-thermalization of some, and maybe all, integrable models and of some non-integrable systems of finite size, such as the Bose Hubbard Model. Our results reveal the crucial importance of finite size effects in the study of thermalization and allow us to point out two alternative routes for thermalization of quantum systems, as well as ways to prune down the correct one.\\
We consider the general situation where a system starts evolving at time $t=0$ from a density matrix $\hat \rho^0$. 

The following time-evolution of any observable $\mathcal{O}$ can be expressed as

$
\aver{\mathcal{O}}(t)=\sum_{\alpha,\beta} \rho^0_{\alpha\beta} e^{-it(E_\alpha-E_\beta)}\langle \beta | \mathcal{O} | \alpha \rangle.
$

Here  $\ket{\alpha}$ are the eigenvectors of the Hamiltonian with corresponding eigenvalues $E_\alpha$ (we use $\hbar=1$). In order to be concrete we will focus on the experimentally relevant situation of a quantum quench, which corresponds to a sudden parameter change of the Hamiltonian at time $t=0$ for a system that is in the ground state for $t<0$. In this case, $\rho_{\alpha\beta}=c_\alpha c^*_\beta$, where $c_\alpha=\langle \alpha|\psi_0\rangle$ is the overlap between the eigenstate $\ket{\alpha}$ of the Hamiltonian after the quench and the ground state $\ket{\psi_0}$ of the  Hamiltonian before the quench ($t=0^-$). Our results can be generalized straightforwardly to a general $\rho^0$.
The typical time behavior of $\aver{\mathcal{O}}(t)$ consists in damped or overdamped oscillations that converge towards a constant average value at long times.
Assuming no degeneracy in eigenenergies, the long-time value 
of $\aver{\mathcal{O}}(t)$  can be computed using the time averaged density matrix, 
$\rho=\sum_\alpha |c_\alpha|^2 |\alpha\rangle\langle\alpha|$~\cite{VonNeumann1929}~\cite{footnote3}. Following Ref.~\cite{RigolOlshanii2008} we call "diagonal ensemble averages"  all averages with respect to
$\rho$ and we use $\aver{\mathcal{O}}_D= \mathrm{Tr}(\rho \mathcal{O})=\sum_\alpha \mathcal{O}_\alpha  |c_\alpha|^2$ with $\mathcal{O}_\alpha= \bra{\alpha}\mathcal{O}\ket{\alpha}$.
An important property of the diagonal ensemble is that under very general conditions \cite{RigolOlshanii2008} the energy per particle has vanishing fluctuations:
\begin{equation}
\label{eq:olshanii-property}
\Delta e:=\frac{ \sqrt{\langle E^2 \rangle_D-\langle E \rangle_D^2}}{L}\rightarrow 0 \quad \textrm{for} \quad L\rightarrow \infty.
\end{equation}
Here $L$ denotes the number of sites and the thermodynamic limit is taken at constant particle density $N/L$. 
Property (\ref{eq:olshanii-property}) means that the distribution of intensive eigenenergies with weights $|c_\alpha|^2$ is peaked for large system sizes.  \\
{ As already anticipated in the introduction the {\it 'eigenstate thermalization hypothesis'} says for generic non-integrable interacting many body systems that the matrix elements $\mathcal{O}_\alpha$ of a few body observable with respect to any eigenstate $\ket{\alpha}$ with eigenenergy $E_\alpha$ equals the microcanonical ensemble average taken at that energy $E_\alpha$. This was first conjectured based on studies of semiclassical systems \cite{Deutsch1991,Srednicki1994} and recently shown numerically to hold for a specific non-integrable system of finite size \cite{RigolOlshanii2008}. 
Were this hypothesis true, an immediate consequence of property (\ref{eq:olshanii-property})  would be that 
averages in the diagonal ensembles coincide with averages in the microcanonical
ensemble at the same energy per particle. This was the explanation of thermalization given for generic non-integrable systems and demonstrated for a specific example \cite{RigolOlshanii2008}. In contrast a finite width 
distribution for specific observables was found numerically 
for a finite size integrable system and claimed to be at the origin of the 
absence of thermalization for this model.}\\
Note, however, that for a finite system there are always finite fluctuations of $\mathcal{O}_\alpha$, whether the system is integrable or not. It follows that a precise characterization of ETH has to involve statements about the evolution of the distribution of $\mathcal{O}_\alpha$ upon approaching the thermodynamic limit. One can prove that generically the width of the distribution of $\mathcal{O}_\alpha$ vanishes in the thermodynamic limit:
\begin{equation}
\label{eq:var}
(\Delta \mathcal{O}_e)^2=\frac{\sum_e \mathcal{O}_\alpha^2 }{\sum_e}-
\left(\frac{\sum_e\mathcal{O}_\alpha}{\sum_e}\right)^2 \rightarrow 0 \; \textrm{for} \; L\rightarrow \infty.
\end{equation}
where $\mathcal{O}$ is an intensive local few body Hermitian operator (or observable); the sum $\sum_e$ is taken over eigenstates $\ket{\alpha}$ with eigenenergies $E_\alpha/L\in [ e-\epsilon;e+\epsilon ]$ where $e$ is the considered energy per particle and $\epsilon$ is a small number that can be taken to zero after the thermodynamic limit. 
Our detailed proof presented in \cite{EPAPS} is based on the vanishing of the fluctuations in the microcanonical ensemble \cite{footnote4}. 
Note that Eq.~(\ref{eq:var}) implies that the fraction of states characterized by a value of $\mathcal{O}_\alpha$ different from the microcanonical average vanishes in the thermodynamic limit.
However, states with different values $\mathcal{O}_\alpha$ may and actually do exist, as we shall show in the following in concrete examples. They are just rare compared to the other ones.
This is not a minor fact since if the $|c_\alpha|^2$s distribution gives an important weight to these rare states, the diagonal ensemble averages will be different from the micro-canonical
one. They keep a memory of the initial state.  
As a consequence, an interpretation of ETH stating that the {\it fraction} of thermal states has to vanish would not guarantee thermalization. Instead the stronger interpretation of ETH,  
stating that the {\it support} of the distribution of the $\mathcal{O}_\alpha$
shrinks around the thermal microcanonical value in the thermodynamic limit, does so because states 
leading to non-thermal averages disappear.  In the following we shall show, in concrete examples, that these rare states indeed do exist and prevent thermalization 
in some integrable infinite systems and in some {\it finite size} non-integrable models, such as the Bose Hubbard one.\\
Our first example is a chain of $L$ harmonic with a mass $m$ and coupling strength $\omega$ described by
$$
H=\frac 1 2 \sum_x\left[ \pi_x^2+m^2\phi_x^2+\sum_{y=\pm 1}\omega^2(\phi_{x+y}-\phi_x)^2\right].
$$
We assume periodic boundary conditions and the usual commutation relations between the operators $\pi_x$ and $\phi_y$ given by $[\phi_x,\pi_y]=i\delta_{x,y}$. Using a suitable standard transformation one can rewrite the Hamiltonian as $H=\sum_{k=0}^{(L-1)/2}\Omega_k (R^\dagger_k R_k+I^\dagger_kI_k)$ with the new creation and annihilation operators $R_k,R_k^\dagger$ and $I_k,I_k^\dagger$ and $\Omega_k^2=m^2+2\omega^2(1-\cos(2\pi k/L))$. 
As a consequence the eigenstates of the Hamiltonian at $t=0^+$ are  
characterized by occupation numbers $\{n^I_k\},\{n^R_k\}$ for the $I$ and $R$ operators.
Following Calabrese and Cardy \cite{CalabreseCardy2007}, we consider now a quantum quench where the system is in the ground state at a certain initial value of $m=m_{i}$ that we switch instantaneously to the final value $m_f$, i.e.~$\Omega_k^{i}\rightarrow \Omega_k^{f}$. 
We focus on the coupling between next-nearest neighbour R-oscillators which reads $\mathcal{G}_2=\frac{1}{L} \sum_k g(k) R^\dagger_k R_k$ with $g(k)=\cos(4 \pi k/L)$ \cite{footnote5}.  
The diagonal matrix element for a state $\alpha=\{n^I_k,n^R_k\}$ is $(\mathcal{G}_2) _\alpha =\frac{1}{L}\sum_k g(k) n_k^R$.
In the large system size limit the number of eigenstates with $(\mathcal{G}_2)_\alpha$ and $E_\alpha/L$ respectively between $\mathcal{G}_2$ and $\mathcal{G}_2+\mathrm{d}\mathcal{G}_2$ and $e$ and $e+\mathrm{d}e$ has the form of a large deviation function, i.e. it is proportional to $\exp(L S_e(\mathcal{G}_2))\mathrm{d}e\mathrm{d}\mathcal{G}_2$, (cf.~\cite{EPAPS}). Physically $S_e$ is just related to the entropy of the system with intensive energy $e$ and an average coupling between next-nearest neighbour equal to $\mathcal{G}_2$. Thus the distribution of $\mathcal{G}_2$ is strongly peaked around the maximum of $S_e(\mathcal{G}_2)$ and has a width of the order $1/\sqrt L$, but its tails extend to non thermal values of $\mathcal{G}_2$. Therefore, this is indeed a case where the width of the distribution of the matrix elements vanishes but the support does not due to the existence of rare states.
Additionally, all the weights $|c_\alpha|^2$ can be computed exactly \cite{EPAPS}.
Their typical value is exponentially small in the size of the system. Thus they can bias significantly the micro-canonical ensemble distribution \cite{footnote6} by counterbalancing the 
difference in cardinality between rare and typical states, which is also exponential in the system size.  
This is indeed what happens as it can be explicitly checked by computing the average value of $\mathcal{G}_2$ in the diagonal ensemble:  $\aver{\mathcal{G}_2}_D=\frac{1}{N}\sum_k f(k) \langle n_k \rangle_{D}$. We find that the distributions of $(\mathcal{G}_2)_\alpha$ in the micro-canonical and diagonal ensembles become infinitely peaked but around two different values, in agreement with \cite{CalabreseCardy2007}, thus explaining the absence of thermalization 
in this model (see \cite{EPAPS} for details).\\
The other example we discuss is the one-dimensional Bose-Hubbard model with one particle per site:
\begin{equation*}
H= - \sum_{j} J\left(b_j^\dagger b^{\phantom{\dagger}}_{j+1}+h.c.\right) + \frac{U}{2} \sum _{j} \hat{n}_j ( \hat{n}_j-1),
\end{equation*}
where  $b^\dagger_j$ and $b_j$ are the bosonic
creation and annihilation operators, and $ \hat{n}_j= b^\dagger_j
b^{\phantom{\dagger}}_j$ the number operators on site $j$. For most values of $U$ and $J$, this model has been shown to be non-integrable \cite{KolovskyBuchleitner2004}. Only in special points, e.g.~($U=0$) and ($J=0$), this model is integrable. 
The first case we consider is a quench from the superfluid state $U_i/J=2$ to $U_f/J=10$. For this quench a non-thermal steady state has been found for long-times~\cite{KollathAltman2007, Roux2009}. The correlations $(\mathcal{G}_1)_\alpha=\sum_j \bra{\alpha}b^\dagger_{j}b_{j+1}\ket{\alpha}/L$ in this non-thermal state (system sizes up to $L=100$, solid horizontal line) do agree well with their diagonal ensemble average ($L=11$, dashed horizontal line), but not with the microcanonical distribution (shaded region).
In this non-integrable situation it is more difficult to disentangle the role of rare states and finite size effects in 
the formation of a non-thermal state as we show in the following.
\begin{figure}
\centerline{\includegraphics[width=\linewidth]{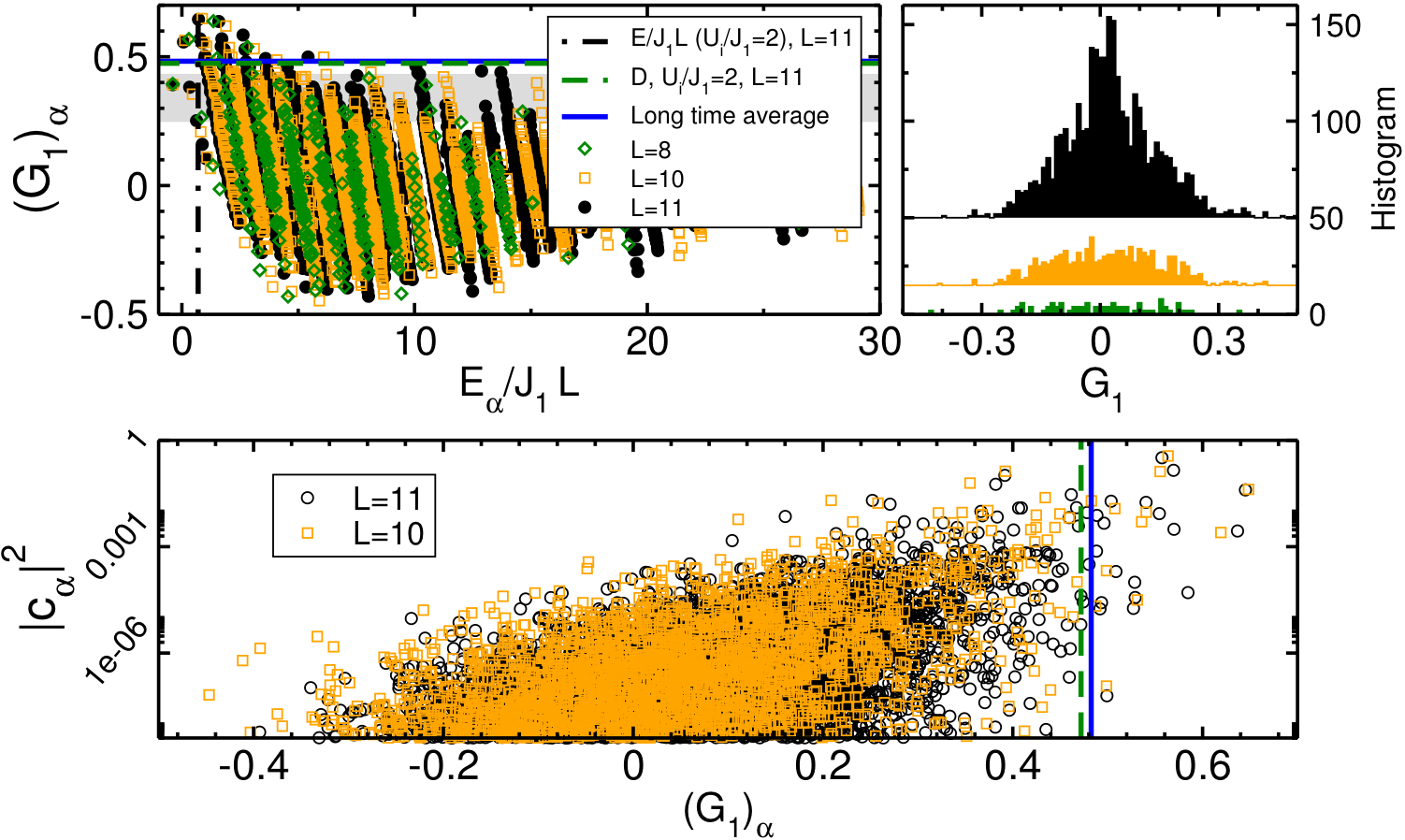}}
\caption{(Color online)
Full diagonalization results of $(\mathcal{G}_1)_\alpha$ versus energy $E_\alpha$ in the even parity, $k=0$ momentum sector for the final Hamiltonian characterized by $U_f/J=10$ (upper panels) and correlation of  $\norm{c_\alpha}$ versus $(\mathcal{G}_1)_\alpha$ for a quench from $U_i/J=2$ (lower panel). Additionally the average energy (dashed-dotted line) after the quench 
and the average value of $\mathcal{G}_1$ obtained from the t-DMRG time-evolution $(L=100)$ (solid line) and the diagonal ensemble for $L=11$ (dashed line) are shown. The shaded region corresponds to the microcanonical average \cite{footnote7}.
The t-DMRG calculations are performed as detailed in Ref.~\cite{KollathAltman2007}. Upper right panel: Distribution of values $\mathcal{G}_1$  for the kinetic energy the values  $E_\alpha/LJ \in [4.5;5.5]$. The average value is removed from the distribution and the histograms are shifted vertically for visibility.
\label{fig:distrU10}}
\end{figure}
\begin{figure}
\centerline{\includegraphics[width=0.9\linewidth]{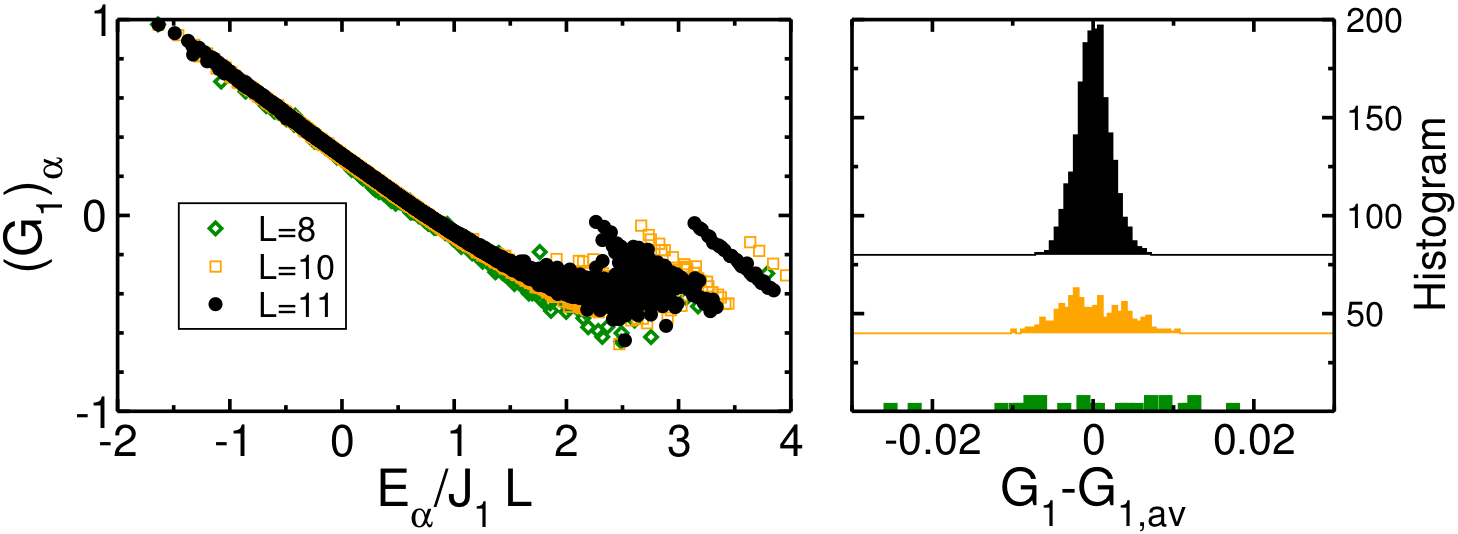}}
\caption{(Color online)
Left panel: full diagonalization results of $(\mathcal{G}_1)_\alpha$ versus energy $E_\alpha$ in the even parity, $k=0$ momentum sector for the final Hamiltonian characterized by $U_f/J=1$. Right panel: distribution of values $\mathcal{G}_1$  for the kinetic energy for values between $E_\alpha/LJ \in [-0.1,0.1]$. The linear trend of the observable in the considered window is removed from the distribution and the histograms are shifted vertically for visibility.
\label{fig:distrU1}}
\end{figure}
First, let us start to consider the validity of eq. (\ref{eq:var}). In Fig.~\ref{fig:distrU10} (upper right panel) we show the correlations $(\mathcal{G}_1)_\alpha$ versus energy $E_\alpha/L$. At low energies an (overlapping) bandstructure is seen. The center of the bands are separated by the interaction energy $U/L$ and have a width proportional to $J/L$.
Within these low energy bands $(\mathcal{G}_1)_\alpha$s decay almost linearly. For intermediate energies a mixing of these energy bands starts to show up (cf.~Fig.~\ref{fig:distrU10} upper right panel $E_\alpha/L\approx 5$) which is weak for small systems and becomes stronger for larger system sizes (cf.~already $L=11$). In most fixed energy intervals the values of the correlations $(\mathcal{G}_1)_\alpha$ are spread considerably. In the upper-right panel the predicted narrowing of the half width of the distribution with increasing system size is clearly visible. In contrast the support does not seem to shrink which might point towards the existence of rare states. This is further supported in  the lower panel of Fig.~\ref{fig:distrU10}, where the weight of the initial state on the final eigenstates is strongly correlated with the values of the $(\mathcal{G}_1)_\alpha$. The weights are much larger for larger values of $(\mathcal{G}_1)_\alpha$, which correspond to the lower energy band edges \cite{Roux2009} and are 
larger than the microcanonical average (shaded region in Fig.~\ref{fig:distrU10}). A general decay of the weights towards lower values of the correlations is evident. This shows that the states which are important for the diagonal ensemble average do not have the microcanonical expectation value, i.e. rare states matter. 
Let us note that for the shown system sizes the average energy after the quench (marked by a vertical line) lies still within the lowest few energy bands. However, we estimated that for the largest sizes considered by t-DMRG ($L=100$), for which there is still no thermalization at accessible time-scales,  the eigenstates with considerable weight will be spread over tens of energy bands and the level statistics close to GOE \cite{KollathLaeuchli2010}. 
We have also studied cases which should be easier from the point of view of thermalization since they are not close to
an integrable point, in particular we discuss $U_f/J=1$ (Fig.~\ref{fig:distrU1}). In this case the distribution of $(\mathcal{G}_1)_\alpha$ is much more peaked 
than for $U_f/J=10$ and its width decreases when increasing system sizes. Additionally, the support of the distribution 
seems to decrease pointing towards thermalization.
Certainly larger system sizes are needed to make any firm statement. \\
As a conclusion, we find that the absence of thermalization for finite size systems can be attributed to two sources: (a) the distribution of the weights $\norm{c_\alpha}^2$ versus energy $E_\alpha$ and the distribution of $\mathcal{O}_\alpha$ in a restricted energy interval may be very broad for finite size systems and (b) states  characterized by a value of $\mathcal{O}_\alpha$ different from the micro-canonical value may have a considerable weight $|c_\alpha|^2$. 
All these phenomena clearly are at play for the finite size Bose Hubbard model investigated above. 
Eq. (\ref{eq:var}) and property (\ref{eq:olshanii-property}) assure that the first origin of non-thermalization will be cured for large enough systems--- the distributions will eventually become infinitely peaked---but not necessarily the second one. Indeed we showed that in some integrable models the origin of non-thermalization stems from the existence of non-thermal eigenstates which are less numerous compared to the thermal ones, but still exist and possibly bias a lot the diagonal expectation values. What happens for non-integrable systems and what is the correct requirement on the $|c_\alpha|^2$s in order to have thermalization in the thermodynamic limit is an open question. Our results
reveal two possible options to obtain thermalization: (1) the support of the distribution of $\mathcal{O}_\alpha$ around the thermal value shrinks to zero in the thermodynamic limit, i.e. rare non-thermal states disappear altogether (as at $U/J=1$ seemingly), (2) rare states exist but the $|c_\alpha|^2$s do not bias too much the micro-canonical distribution toward them. Since the only apriori distinction between rare and typical states is that the latter are overwhelming more numerous, a plausible (but not necessary) assumption, leading to thermalization is that the $|c_\alpha|^2$s sample rather uniformly states with the same energy. Note that the existence of rare states for very large non-integrable models is not completely unreasonable as suggested by mathematical physics results obtained in the semi-classical limit \cite{FaureBievre2003,Nonnenmacher2008}.
 Both scenarios are testable in numerical experiments. One has to study 
how the support of the distribution of $\mathcal{O}_\alpha$ evolves with the size of the system to understand whether (1) is realized, see \cite{EPAPS} for a first attempt. 
In order to study (2), one can use the (von Neumann) Kullback-Leibler (KL) entropy $S_{KL}$ \cite{CoverThomas2006} of the Gibbs distribution with respect to the diagonal ensemble \cite{footnote}. A 'rather uniform sampling' would correspond
to a zero intensive $S_{KL}$ in the thermodynamic limit. \\
We conclude stressing that thermalization after a quantum quench appears to be a property that emerge for large enough system sizes. Understanding the physics behind this "finite size thermalization length" and its dependence on the distance from integrability is a very interesting problem worth investigating in the future, specially because some cold atomic systems  
may well be below this thermalization threshold.\\
We would like to thank P. Calabrese, D. Huse, S. Kehrein, M. Olshanii, P. Reimann, M. Rigol, and G. Roux for fruitful discussions. This
work was partly supported by the 'Triangle de la Physique', DARPA-OLE and the ANR ('FAMOUS').


\newpage
\includepdf[pages={1}]{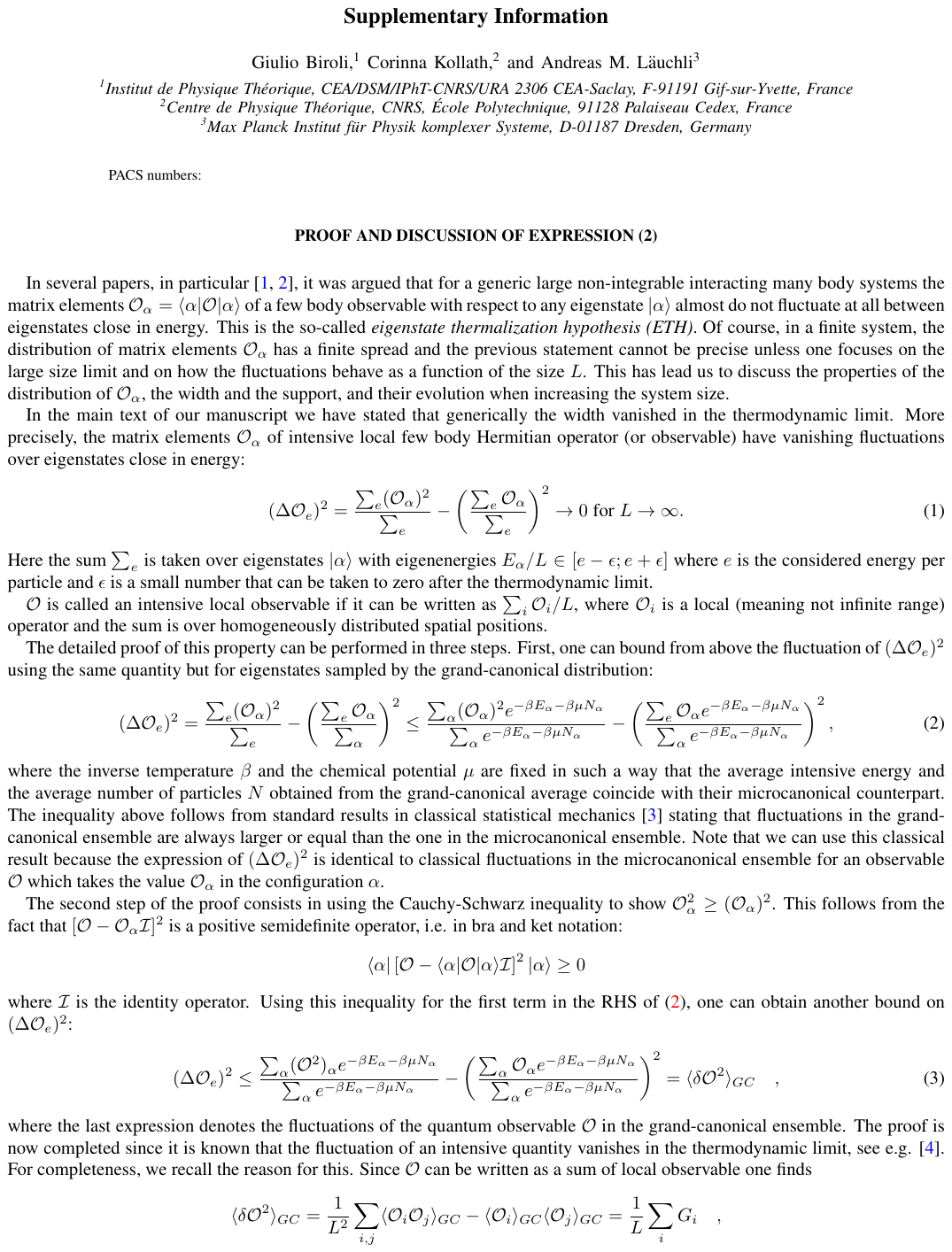}
$\mbox{}$
\newpage
\includepdf[pages={2}]{epaps_arxiv.pdf}
$\mbox{}$
\newpage
\includepdf[pages={3}]{epaps_arxiv.pdf}
$\mbox{}$
\newpage
\includepdf[pages={4}]{epaps_arxiv.pdf}
$\mbox{}$
\newpage
\includepdf[pages={5}]{epaps_arxiv.pdf}
$\mbox{}$
\newpage
\includepdf[pages={6}]{epaps_arxiv.pdf}
$\mbox{}$
\newpage
\includepdf[pages={7}]{epaps_arxiv.pdf}
$\mbox{}$
\newpage


\section*{Supplementary material}

# Supplementary Information

Giulio Biroli,[1] Corinna Kollath,[2] and Andreas M. Läuchli[3]

[1]*Institut de Physique Théorique, CEA/DSM/IPhT-CNRS/URA 2306 CEA-Saclay, F-91191 Gif-sur-Yvette, France*
[2]*Centre de Physique Théorique, CNRS, École Polytechnique, 91128 Palaiseau Cedex, France*
[3]*Max Planck Institut für Physik komplexer Systeme, D-01187 Dresden, Germany*

PACS numbers:

## PROOF AND DISCUSSION OF EXPRESSION (2)

In several papers, in particular [1, 2], it was argued that for a generic large non-integrable interacting many body systems the matrix elements $\mathcal{O}_\alpha = \langle\alpha|\mathcal{O}|\alpha\rangle$ of a few body observable with respect to any eigenstate $|\alpha\rangle$ almost do not fluctuate at all between eigenstates close in energy. This is the so-called *eigenstate thermalization hypothesis (ETH)*. Of course, in a finite system, the distribution of matrix elements $\mathcal{O}_\alpha$ has a finite spread and the previous statement cannot be precise unless one focuses on the large size limit and on how the fluctuations behave as a function of the size $L$. This has lead us to discuss the properties of the distribution of $\mathcal{O}_\alpha$, the width and the support, and their evolution when increasing the system size.

In the main text of our manuscript we have stated that generically the width vanished in the thermodynamic limit. More precisely, the matrix elements $\mathcal{O}_\alpha$ of intensive local few body Hermitian operator (or observable) have vanishing fluctuations over eigenstates close in energy:

$$(\Delta\mathcal{O}_e)^2 = \frac{\sum_e (\mathcal{O}_\alpha)^2}{\sum_e} - \left(\frac{\sum_e \mathcal{O}_\alpha}{\sum_e}\right)^2 \to 0 \text{ for } L \to \infty. \tag{1}$$

Here the sum $\sum_e$ is taken over eigenstates $|\alpha\rangle$ with eigenenergies $E_\alpha/L \in [e-\epsilon; e+\epsilon]$ where $e$ is the considered energy per particle and $\epsilon$ is a small number that can be taken to zero after the thermodynamic limit.

$\mathcal{O}$ is called an intensive local observable if it can be written as $\sum_i \mathcal{O}_i/L$, where $\mathcal{O}_i$ is a local (meaning not infinite range) operator and the sum is over homogeneously distributed spatial positions.

The detailed proof of this property can be performed in three steps. First, one can bound from above the fluctuation of $(\Delta\mathcal{O}_e)^2$ using the same quantity but for eigenstates sampled by the grand-canonical distribution:

$$(\Delta\mathcal{O}_e)^2 = \frac{\sum_e (\mathcal{O}_\alpha)^2}{\sum_e} - \left(\frac{\sum_e \mathcal{O}_\alpha}{\sum_\alpha}\right)^2 \leq \frac{\sum_\alpha (\mathcal{O}_\alpha)^2 e^{-\beta E_\alpha - \beta\mu N_\alpha}}{\sum_\alpha e^{-\beta E_\alpha - \beta\mu N_\alpha}} - \left(\frac{\sum_e \mathcal{O}_\alpha e^{-\beta E_\alpha - \beta\mu N_\alpha}}{\sum_\alpha e^{-\beta E_\alpha - \beta\mu N_\alpha}}\right)^2, \tag{2}$$

where the inverse temperature $\beta$ and the chemical potential $\mu$ are fixed in such a way that the average intensive energy and the average number of particles $N$ obtained from the grand-canonical average coincide with their microcanonical counterpart. The inequality above follows from standard results in classical statistical mechanics [3] stating that fluctuations in the grand-canonical ensemble are always larger or equal than the one in the microcanonical ensemble. Note that we can use this classical result because the expression of $(\Delta\mathcal{O}_e)^2$ is identical to classical fluctuations in the microcanonical ensemble for an observable $\mathcal{O}$ which takes the value $\mathcal{O}_\alpha$ in the configuration $\alpha$.

The second step of the proof consists in using the Cauchy-Schwarz inequality to show $\mathcal{O}_\alpha^2 \geq (\mathcal{O}_\alpha)^2$. This follows from the fact that $[\mathcal{O} - \mathcal{O}_\alpha \mathcal{I}]^2$ is a positive semidefinite operator, i.e. in bra and ket notation:

$$\langle\alpha|\left[\mathcal{O} - \langle\alpha|\mathcal{O}|\alpha\rangle\mathcal{I}\right]^2|\alpha\rangle \geq 0$$

where $\mathcal{I}$ is the identity operator. Using this inequality for the first term in the RHS of (2), one can obtain another bound on $(\Delta\mathcal{O}_e)^2$:

$$(\Delta\mathcal{O}_e)^2 \leq \frac{\sum_\alpha (\mathcal{O}^2)_\alpha e^{-\beta E_\alpha - \beta\mu N_\alpha}}{\sum_\alpha e^{-\beta E_\alpha - \beta\mu N_\alpha}} - \left(\frac{\sum_\alpha \mathcal{O}_\alpha e^{-\beta E_\alpha - \beta\mu N_\alpha}}{\sum_\alpha e^{-\beta E_\alpha - \beta\mu N_\alpha}}\right)^2 = \langle\delta\mathcal{O}^2\rangle_{GC} \quad, \tag{3}$$

where the last expression denotes the fluctuations of the quantum observable $\mathcal{O}$ in the grand-canonical ensemble. The proof is now completed since it is known that the fluctuation of an intensive quantity vanishes in the thermodynamic limit, see e.g. [4]. For completeness, we recall the reason for this. Since $\mathcal{O}$ can be written as a sum of local observable one finds

$$\langle\delta\mathcal{O}^2\rangle_{GC} = \frac{1}{L^2}\sum_{i,j}\langle\mathcal{O}_i\mathcal{O}_j\rangle_{GC} - \langle\mathcal{O}_i\rangle_{GC}\langle\mathcal{O}_j\rangle_{GC} = \frac{1}{L}\sum_i G_i \quad,$$

where $G_{i-j} = \langle \mathcal{O}_i \mathcal{O}_j \rangle_{GC} - \langle \mathcal{O}_i \rangle_{GC} \langle \mathcal{O}_j \rangle_{GC}$ is the equilibrium correlation function at distance $i-j$ and we have used explicitly translation invariance (quenched disordered systems are out of the scope of this article). Thus, we have found that as long as $\sum_i G_i$ is finite or, at least increase with $L$ with a power less than one:

$$(\Delta \mathcal{O}_e)^2 \leq \frac{1}{L} \sum_i G_i \to 0 \text{ for } L \to \infty.$$

The above property on $\sum_i G_i$ is true for all quantum statistical mechanics models except pathological ones. In fact, were this property not verified, intensive observables would fluctuate and equilibrium thermodynamics would be ill-defined. In the majority of cases $G_i$ decreases fast enough (actually exponentially) so that the sum on $i$ is finite. This leads to $(\Delta \mathcal{O}_e)^2$ going to zero at least as fast as $1/L$. In some cases $G_i$ decreases as a power law [5] with a power such that the sum is not convergent. In this case $\sum_i G_i \propto L^\alpha$ with $\alpha < 1$ and so $(\Delta \mathcal{O}_e)^2$ goes to zero at least as fast as $1/L^{1-\alpha}$.

Let us now discuss two subtleties. The first one concerns the hypothesis that $\mathcal{O}$ is an intensive local observable. Actually, we do not need to assume that $O$ is an intensive observable if translation invariance holds both for the Hamiltonian and the initial condition $|\psi_0\rangle$, as it is the case for the models considered in this work. The reason is that in this case one can restrict all the analysis directly to translation invariant states $|\alpha\rangle$ since all the other ones have a zero overlap with $|\psi_0\rangle$. As a consequence, $\sum_i \langle \alpha | \mathcal{O}_i | \alpha \rangle / L = \langle \alpha | \mathcal{O}_j | \alpha \rangle$ for any $j$ and there is no difference between local and intensive observables.

Finally, the other subtlety we would like to discuss is whether our discussion applies to the momentum distribution, which at least at first sight looks like a non-local quantity. The answer is yes only if we do not require, rigorously speaking, an infinite precision on the momentum distribution. Let us explain this subtle point in detail. We define the coarse grained momentum distribution $\hat{G}_c$ as:

$$\hat{G}_c(k) = \int dk' \frac{\exp\left(-\frac{(k-k')^2}{2\Delta}\right)}{\sqrt{2\pi\Delta}} \hat{G}(k')$$

where $\hat{G}(k)$ is the momentum distribution, i.e. the Fourier transform of $G(x)$, and $\Delta$ is a coarse graining width, which is arbitrary small but finite in the large $L$ limit. Note also that for simplicity we consider the system in 1D and on the continuum, i.e. sums are replaced by integrals. Eq. (1) applies to $\hat{G}_c(k)$ and not to $\hat{G}(k)$ because the former is a local observable, whereas the latter is not. One easy way to see that consists in writing the Fourier transform of $\hat{G}_c(k)$:

$$\int \frac{dk}{2\pi} e^{ikx} \hat{G}_c(k) = G(x) \exp\left(-\Delta \frac{x^2}{2}\right)$$

We find that $\hat{G}_c(k)$ is, roughly speaking, the Fourier transform of $G(x)$ truncated for values of $x \gg 1/\sqrt{\Delta}$. Thus, it is a local quantity. Instead $\hat{G}_c(k)$ is not as it can be readily checked by considering its $k=0$ value.

## HARMONIC OSCILLATOR CHAIN

### Detailed derivation of $c_\alpha$ and comparison to GGE

In the following we provide some details on the computation performed for the Harmonic oscillator chain of length $L$. We recall that the Hamiltonian is

$$H = \frac{1}{2} \sum_x \left[ \pi_x^2 + m^2 \phi_x^2 + \sum_{y=\pm 1} \omega^2 (\phi_{x+y} - \phi_x)^2 \right].$$

By going to Fourier space one can decouple the oscillators. Defining

$$\phi_k^R = \frac{1}{\sqrt{2L}} \sum_{x=0}^{L-1} \cos(2\pi xk/L) \phi_x \qquad \pi_k^R = \frac{1}{\sqrt{2L}} \sum_{x=0}^{L-1} \cos(2\pi ixk/L) \pi_x$$

$$\phi_k^I = \frac{1}{\sqrt{2L}} \sum_{x=0}^{L-1} \sin(2\pi xk/L) \phi_x \qquad \pi_k^I = \frac{1}{\sqrt{2L}} \sum_{x=0}^{L-1} \sin(2\pi ixk/L) \pi_x$$

we find the new expression for the Hamiltonian:

$$H = \frac{1}{2} \sum_{k=0}^{(L-1)/2} \sum_{i=R,I} \left[ (\pi_k^i)^2 + \Omega_k^2 (\phi_k^i)^2 \right],$$

where $\Omega_k = m^2 + 2\omega^2(1 - \cos(2\pi k/L))$ [14].

The initial wave function, which we will denote by $\prod_{k=1}^{L/2} |n_k^R = 0>_i |n_k^I = 0>_i$, reads:

$$\prod_{k=1}^{L/2} \frac{e^{-\Omega_k^i (\phi_k^I)^2/2}}{(\pi/\Omega_k^i)^{1/4}} \frac{e^{-\Omega_k^i (\phi_k^R)^2/2}}{(\pi/\Omega_k^i)^{1/4}}$$

where $\Omega_k^i$ is the value of $\Omega_k$ just before the quench, its value just after will be denoted $\Omega_k^f$. The quantum quench changes only the mass, therefore the scalar product $c_\alpha == \langle \alpha | \psi_0 \rangle$ is given by

$$\prod_{k=1}^{L/2} <n_k^I|n_k=0>_i <n_k^R|n_k=0>_i \quad ,$$

Thus, to obtain $c_\alpha$ one has to compute one scalar product between a generic eigenstate of the harmonic oscillator with frequency $\Omega_k^f$ and the ground state of the harmonic oscillator with frequency $\Omega_k^i$. This reads

$$<n_k|n_k=0>_i = \int dx \frac{e^{-\Omega_k^i x^2/2}}{(\pi/\Omega_k^i)^{1/4}} \frac{e^{-\Omega_k^f x^2/2}}{(\pi/\Omega_k^f)^{1/4}} \frac{H_{n_k}(\sqrt{\Omega_k^f}x)}{\sqrt{n_k! 2^{n_k}}} = \left(\frac{\Omega_k^i}{\Omega_k^f}\right)^{1/4} \frac{1}{\sqrt{A 2^{n_k}}} \frac{\sqrt{n_k!}}{(n_k/2)!} (-1 + 1/A)^{n_k/2} E_{n_k}$$

where $E_{n_k}$ equals one for $n_k$ even and zero otherwise, $A = (1 + \Omega_k^i/\Omega_k^f)/2$ and $H_n$ denotes the Hermite polynomial of degree $n$.

We finally obtain the expression for $|c_\alpha|^2$:

$$|c_\alpha|^2 = \prod_k \left(\frac{\Omega_k^i}{\Omega_k^f}\right) \frac{1}{A 2^{n_k^I}} \frac{n_k^I!}{(n_k^I/2)!^2} (-1 + 1/A)^{n_k^I} E_{n_k^I} \times \frac{1}{A 2^{n_k^R}} \frac{n_k^R!}{(n_k^R/2)!^2} (-1 + 1/A)^{n_k^R} E_{n_k^R}$$

Note that instead the Generalized Gibbs Ensemble (GGE) [6] would have suggested:

$$|<n_k|n_k=0>_i|^2 = (1 - e^{-\beta_k}) e^{-\beta_k n_k}.$$

Clearly, the true distribution is different from the functional point of view:

$$|<n_k|n_k=0>_i|^2 = \left(\frac{\Omega_k^i}{\Omega_k^f}\right)^{1/2} \frac{1}{A 2^{n_k}} \frac{n_k!}{(n_k/2)!^2} (-1 + 1/A)^{n_k} E_{n_k}$$

Note that in the GGE approach the value of the generalized temperature $\beta_k$ is fixed by requiring that the average value of $n_k$ is equal to its $t = 0$ value. This is nothing else than the average of $n_k$ with the true distribution we computed. So by definition the two distributions have the same first moment, which is

$$<n_k> = \sum_{n_k} n_k |<n_k|n_k=0>_i|^2 = \frac{(A-1)^2}{2(2A-1)} = \frac{1}{4}\left(\frac{\Omega_k^i}{\Omega_k^f} + \frac{\Omega_k^f}{\Omega_k^i}\right) - 1/2 \quad ,$$

in agreement with Calabrese and Cardy [7]. Requiring that the above result equals $1/(e^{\beta_k} - 1)$ we fix the value of $\beta_k$. The second moment of the two distributions, i.e. the average value of the square of the occupation number, instead, is different: the GGE would lead to the result:

$$<n_k^2>_{GGE} = \frac{e^{\beta_k} + 1}{(e^{\beta_k} - 1)^2}$$

The result corresponding to the true distribution (expressed in terms of the $\beta_k$ fixed by imposing the value of $<n_k>$) is:

$$<n_k^2> = \frac{2e^{\beta_k} + 1}{(e^{\beta_k} - 1)^2}$$

Note that this discrepancy does not mean GGE is incorrect as far as local observables are concerned. In fact one can show that all averages of local observables can be rewritten in terms of the two point function for the harmonic oscillator chain. Since the latter is a function of $<n_k>$ only, GGE works. But any other distribution characterized by the same value of $<n_k>$ would too. From this perspective, GGE is not wrong but not useful either. The situation is very different from the usual Gibbs ensemble. In this case, to obtain all local observables one needs to only fix the energy and the particle number.

On the validity of the GGE compare also the references Barthel and Schollwoeck [8] who have chosen as well the harmonic oscillators as one of their example and Eckstein and Kollar [9].

### Large Deviations

In the main text, we focus on the distribution of the coupling between next-nearest neighbour R-oscillators which reads $\mathcal{G}_2 = \frac{1}{L}\sum_k g(k) R_k^\dagger R_k$ with $g(k) = \cos(4\pi k/L)$. The diagonal matrix element for a state $\alpha = \{n_k^I, n_k^R\}$ is simply $(\mathcal{G}_2)_\alpha = \frac{1}{L}\sum_k g(k) n_k^R$. Let us consider the number of eigenstates with $(\mathcal{G}_2)_\alpha$ and $E_\alpha/L$ respectively between $\mathcal{G}_2$ and $\mathcal{G}_2 + \mathrm{d}\mathcal{G}_2$ and $e$ and $e + \mathrm{d}e$:

$$\mathcal{N}(e, \mathcal{G}_2) = \sum_\alpha \delta\left(e - E_\alpha/L\right) \delta\left(\mathcal{G}_2 - \frac{1}{L}\sum_k g(k) n_k\right) \tag{4}$$

It can be easily shown that the scaling with $L$ of $\mathcal{N}(e, \mathcal{G}_2)$ in the large $L$ limit reads:

$$\lim_{L\to\infty} \frac{\log \mathcal{N}(e, \mathcal{G}_2)}{L} = S_e(\mathcal{G}_2).$$

This follows from the fact that the expression above is identical to the one of the entropy of a classical system with configuration $\{\alpha\}$ and classical observables $E_\alpha$ and $\frac{1}{L}\sum_k g(k) n_k$. The outcome is that the distribution of $\mathcal{G}_2$, for a given value of $e$, is strongly peaked around the maximum of $S_e(\mathcal{G}_2)$ with a width of the order $1/\sqrt{L}$. However rare states do exist, but they are exponentially less numerous than the typical ones.

We present now a general argument that is helpful to grasp why large deviations and rare states explain the absence of thermalization for the integrable system we focus on (and others). Let us focus on:

$$Z(e, o) = \sum_\alpha |c_\alpha|^2 \delta\left(e - E_\alpha/L\right) \delta\left(o - \mathcal{O}_\alpha\right) \tag{5}$$

For the same reason discussed above, this scales in the large $L$ limit as $\exp LF(e, o)$ where $F$ is a large deviation function. Note that the expression above is very similar to eq. (4). Were the $|c_\alpha|^2$s uniform on states with intensive energy $e^*$ (equal to the average energy after the quench) then $F(e^*, o)$ would be equal to $S(e^*, o)$ up to a constant. Instead, the existence of rare states and non-uniform $|c_\alpha|^2$s, may lead to a situation where rare states are weighted considerably more than typical ones and, hence, $F(e^*, o)$ is different from $S(e^*, o)$ (beyond the constant term), as it can be shown by an explicit computation for $o = \mathcal{G}_2$ for the Harmonic oscillator chain.

We can now rewrite the average of an observable $\mathcal{O}$ as:

$$\sum_\alpha |c_\alpha|^2 \mathcal{O}_\alpha = \int \mathrm{d}o \mathrm{d}e Z(e, o) o$$

The integral on the RHS is infinitely peaked in the large $L$ limit around the values $e^*, o^*$ maximizing $F(e, o)$. We know that the value $e^*$ coincides with the average energy after the quench because the energy is conserved. However, in cases where $F(e^*, o)$ is different from $S(e^*, o)$ beyond the constant term, the value of $o$ maximizing $F$ is different from the one maximizing $S$. The former is the long-time value of the observable $\mathcal{O}$, whereas the latter corresponds to the microcanonical result. As a conclusion, we find that absence of thermalization is due to the fact that the weights $c_\alpha$ strongly bias the microcanonical distribution, leading to a large deviation function $F$ different from $S$.

### VALIDITY OF ETH IN NON-INTEGRABLE SYSTEMS OF FINITE SIZE

In this section we give more details on the validity of the ETH in finite size non-integrable systems. In particular we consider the distribution of different observables in the Bose-Hubbard model and analyze its dependence on the system parameters as

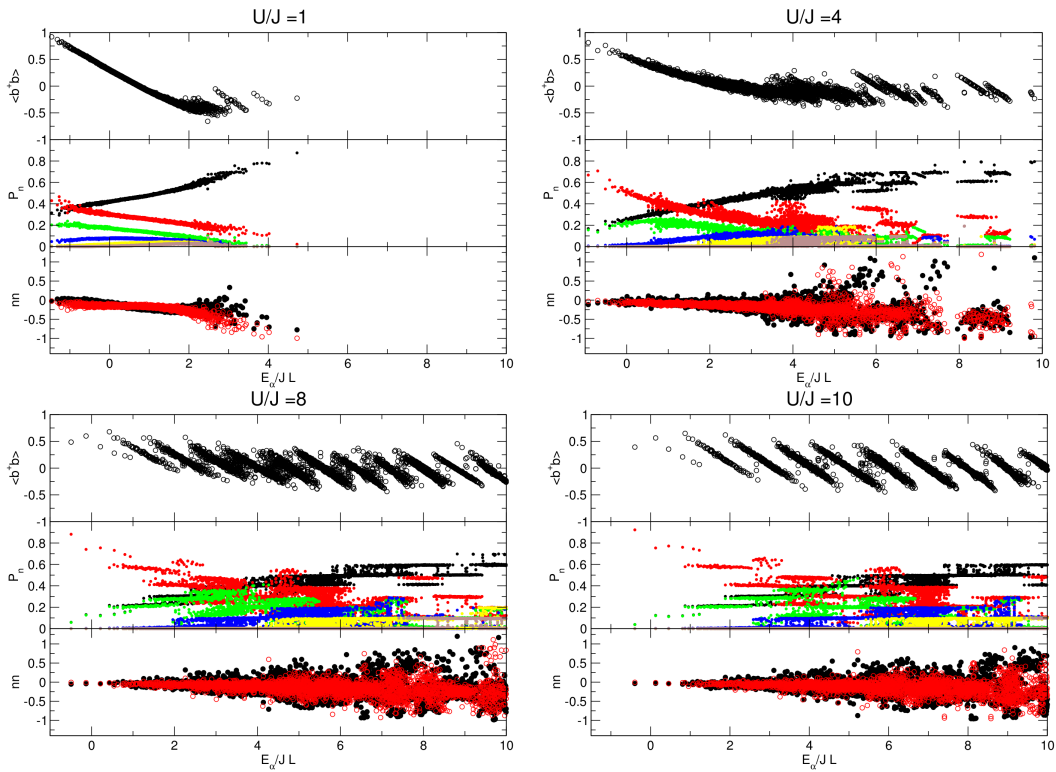

FIG. 1: Expectation values $\mathcal{O}_\alpha$ of the kinetic energy, occupation probability, and the density-density correlations for the eigenvectors $|\alpha\rangle$ at energy $E_\alpha/L$ at the interaction strength $U/J = 1$, $U/J = 4$, $U/J = 8$, and $U/J = 10$ in a system of length $L = 10$. In the central panel the probabilities $\langle P_{\tilde{n}} \rangle_\alpha$ to find $\tilde{n} = 0, 1, 2, 3, 4, 5$ are decoded by the colors black, red, green, blue, yellow, and brown, respectively. In the lowest panel the distances $d = 1$ and $2$ correspond to black and red, respectively.

interaction strength and system length. Similar results for the single particle correlation function and in particular the dependence on the system size have been recently discussed by Roux [10]. However, for completeness we include the results for the single particle correlation function in the present supplementary material. Roux further discussed the momentum distribution whereas we focus on additional real space observables as the probability of finding different occupation on a site or density-density correlations.

In Fig. 1 we show the expectation values of the single-particle correlation $\mathcal{G}_1 = \langle b_i^\dagger b_{i+1} \rangle_\alpha$, the probability of finding different occupancies $\tilde{n}$ on a certain site $\langle P_{\tilde{n}} \rangle_\alpha$, and the density-density correlations $\langle n_i n_{i+d} \rangle_\alpha - \langle n_i \rangle_\alpha \langle n_{i+d} \rangle_\alpha$ for $d = 1$ and $d = 2$ versus the eigenenergy $E_\alpha/L$. These are shown for different values of the interaction strength $U/J$ and a system of length $L = 10$ [15]. Let us first focus on the two extreme cases of weak and strong interaction. At weak interaction (cf. $U/J = 1$) the expectation value distributions for the considered observables are very narrow and only in the upper part of the spectrum a larger spread occurs. The expectation values of the single-particle correlation drop with increasing energy. It starts with a value close to one for the superfluid ground state and goes down to $-0.5$. Low lying collective excitations exist and only a small finite size gap can be seen. Let us note that for the integrable system $U = 0$ the expectation value of the nearest neighbour single particle correlation becomes trivially a single value, since it fully determines the energy. At the same time the probability distribution of the occupancies changes its character with increasing energy. For the low lying states the probability to find 0, 1, and 2 particles on a certain site are very high (it is a Poissonian distribution for the ground state). In contrast at high energies the probability to find no particle at a site is dominating. The same development can be seen in the density-density correlations which is first close to zero and then becomes close to $\pm 1$. For these weak interactions and the considered observables a narrow distribution as predicted by eq. (1) in the thermodynamic limit is already seen for the small system lengths. In Fig. 2 of the main paper we show the distribution of values of the single particle correlations for a certain energy per site for different system lengths. One clearly sees that the support of the distribution is very narrow and that the distribution becomes more and more peaked with increasing system size. The support even seems to shrink increasing the system size. Of course larger system sizes are needed to make any firm statement. Let us note that in the regime of $U \propto J$ and the considered system sizes that spectral properties are chaotic [13].

The behaviour at low interaction strength has to be contrasted with many energy bands at large interaction strength. For

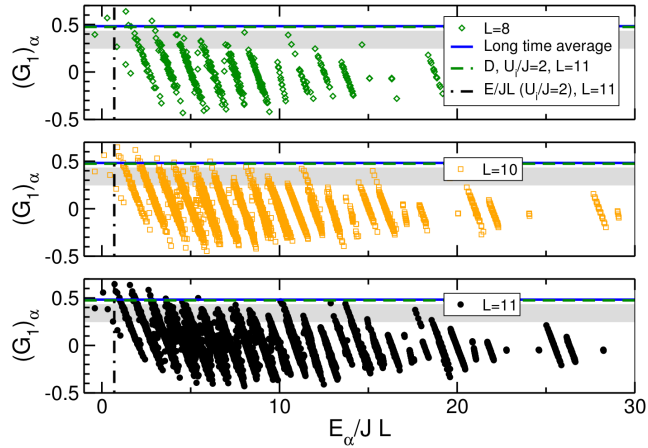

FIG. 2: Expectation values of the kinetic energy $\mathcal{G}_1$ for the eigenvectors $|\alpha\rangle$ at energy $E_\alpha$ at the interaction strength $U/J = 10$ for different system sizes $L = 8$, $L = 10$, and $L = 11$. Additionally the average energy (dashed-dotted line) after the quench and the average value of $\mathcal{G}_1$ obtained from the t-DMRG time-evolution ($L = 100$) (solid line) and the diagonal ensemble for $L = 11$ (dashed line) are shown. The shaded region corresponds to the microcanonical average. The t-DMRG calculations are performed as detailed in Ref. [11].

example at $U/J = 10$ the expectation values show a clear separation into distinguishable energy bands with distance of the order of $U/J_1 L$ and width $\propto J$. The Mott insulating ground state is well separated from the next energy band (the single particle-hole excitation band). It has almost no single particle correlations and the probability to find one particle at a site is very large. As seen in the distribution of the occupation probability the lower bands can be clearly labeled by delocalized Fock-like excitations. For example the third lowest band corresponds to the excitation with 3 particle-hole excitations (relatively wide spread in energy or a single triply occupied site (narrow in energy). For intermediate energies many different excitations exist at the same energy (cf. $E_\alpha/LJ_1 \approx 5$). Within each band the expectation value of the single-particle correlation varies from a positive value at the lower band edge to a negative value at the higher band edge. The density-density correlations show a narrow distribution in the low energy spectrum which becomes considerably spread in the upper part of the spectrum.

The variation with system length is shown in Fig. 2. Whereas for small system length the bands are well separated at larger system sizes in the intermediate energy regime the energy bands become strongly mixed. Let us note that for these system sizes the level statistics resembles already a GOE like distribution, even though it still evolves with system length[12].

In Fig. 1 of the main paper we show the histogram for the distribution of the single particle correlations in an fixed energy density interval. The peaking of the distribution for large system sizes is evident. In the thermodynamic limit we expect that this distribution becomes peaked around a mean value as predicted by eq. (1). However, as for small $U/J$ we do not find any hint that ETH will eventually hold, since the support of the distribution does not show any substantial modification increasing $L$. Of course larger system sizes are needed to make any firm statement. Note that the $J/U = 0$ limit is exactly solvable and corresponds to a classical statistical mechanism model. In this case it is possible to compute the fluctuations of $P_n$ from state to state and check that eq. (1) is verified. Thus, even though, for large $U/J$ there is still a considerable spread in the distribution of $P_n$ in Fig. 1, one can be sure that in the thermodynamic limit the width of the distribution vanishes.

How the expectation values evolve from weak to strong interaction from a very narrow band to the gapped band structure is shown for chosen intermediate values of the interaction strength. Increasing the interaction strength causes the expectation values to develop first the band structure for high energies, while at lower energies the expectation values stay narrowly distributed. At around $U/J = 4$ the arising energy bands can be followed down to low energies. The previously monotoneous decay in the single-particle correlations forms an envelope for the still overlapping energy bands. At $U/J = 8$ already a band structure can be guessed. However, still a lot of values exist which mix the different energy bands.

## Long-time steady state in finite systems

The definition of equilibration in a closed finite system is subtle. In this section we consider the relaxation of different observables to the diagonal values $\langle O \rangle_D$. The observables we will consider are the on-site particle distribution $P_n$ and the single particle correlation functions $\Re \langle b_j^\dagger b_{j+r} \rangle$ for different distances $r$. The on-site particle distribution is directly related to the reduced density matrix of a single site by $\rho_j = \sum_n P_n |n\rangle_j \langle n|_j$ where the state $|n\rangle_j$ is the state of site $j$ with occupation $n$.

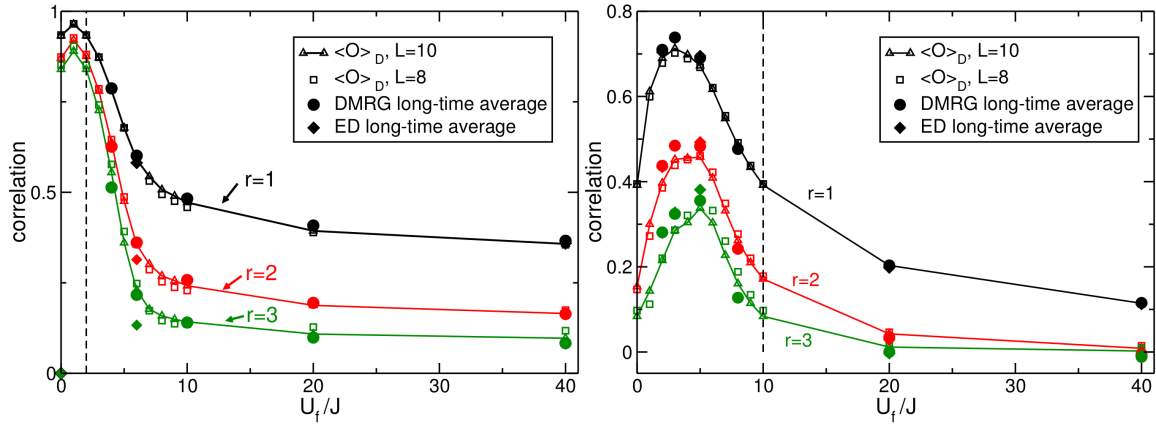

FIG. 3: Comparison of the long time average of the time-evolution calculated with DMRG/ED and the expectation value $\langle b_j^\dagger b_{j+r}\rangle_D$ for two different initial values of $U_i/J = 2$ (upper panel) and $U_i/J = 10$ (lower panel) indicated by the dashed vertical line. Very good agreement between the long-time value and the diagonal term is found even for large values of $U_f/J$.

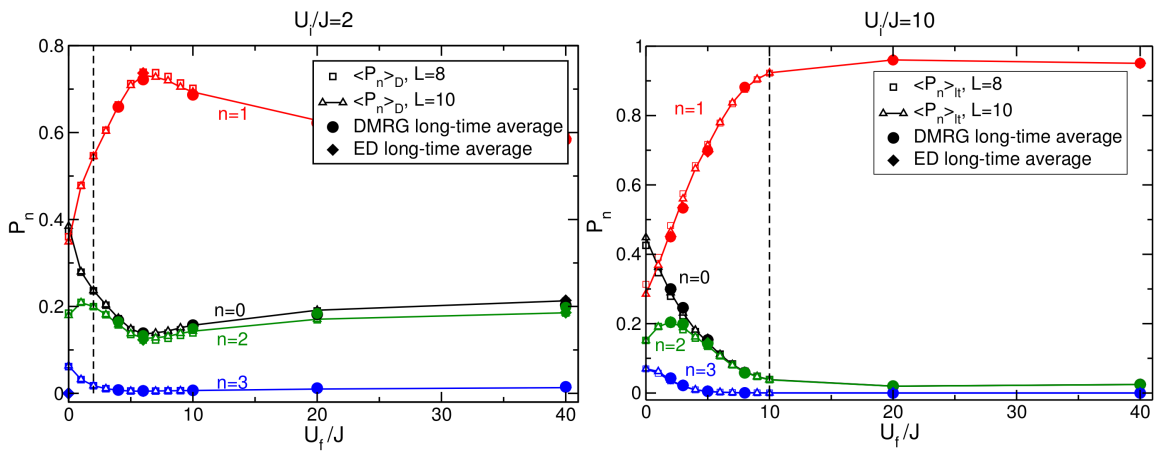

FIG. 4: Comparison of the long time average of the time evolution determined by DMRG/ED and the expectation values $\langle P_n\rangle_D$ for the probability to find $n$ particles on a certain site. Two different initial states are considered, $U_i/J = 2$ (upper panel) and $U_i/J = 10$ (lower panel). Very good agreement between the long-time value and the diagonal term is found for all values of $U_f/J$.

This means looking at the on-site particle distribution gives all information about the local site.

We show that in small systems the considered quantities have reached a quasi-steady state the time-average of which corresponds to the expected diagonal average $\langle O\rangle_D$ to a very good accuracy. We demonstrate this using two different inital states. The first initial state is the superfluid ground state at weak interaction $U_i/J = 2$ and the second is a Mott-insulating state corresponding to the ground state of a system with interaction strength $U_i/J = 10$. The sudden change of the interaction parameter is performed to various final values including both changes to the superfluid phase and to the Mott-insulating regime. Even the real time evolution results for larger system length obtained by t-DMRG agree to a very good extend and only slight finite system effects are visible. This, however, might stem from the fact that only a short time steady state can be reached within these simulations and slow decay processes are not taken into account. In the following we discuss the results for the different observables in more detail.

For all cases, the long-time average value of the time-dependent simulations has been obtained by a fit of a constant value to the long time data[11]. The exact-diagonalization data for the time-evolution is obtained in a system with periodic boundary conditions (L=14, local cutoff=4) whereas the DMRG data is taken from the center of an open system with up to 128 sites and a local cutoff of 6 states. The expectation of the diagonal ensemble is evaluated using exact-diagonalization methods for system length up to 10 sites.

In Fig. 3 the long time average of the single particle correlations are compared to the diagonal expectation $\langle b_j^\dagger b_{j+r}\rangle_D$[16]. For both initial states an overall good agreement between the average long-time limit and the expected value is found for most



\begin{thebibliography}{25}
\bibitem{footnote1}
The coupling to a thermalized bath is not satisfactory to explain thermalization from a fundamental point of 
view because one has then to justify why the reservoir is thermalized in the first place.
\bibitem{footnote2}By thermalization we mean that all the long-time averages of local few body observables coincide with Gibbs averages corresponding to the same intensive energy and particle density.
\bibitem{footnote4} But it is not equivalent to it since the eigenstates $\alpha$ are in general not eigenstates of $\mathcal{O}$.
\bibitem{footnote5} Although $\mathcal{G}_2$ does not evolve with time, one can easily construct dynamical observables, using two point correlation functions of $\phi_x$, whose long-time limit is equal to $\mathcal{G}_2$. Thus, for simplicity, we just focus on $\mathcal{G}_2$.
\bibitem{footnote7} The microcanonical average in these small system is depending on the exact energy interval taken. Therefore several intervals of width $(2,5,10,15,20 )J/L$ have been chosen. The minimum and maximum value obtained define the shaded region.
\end{thebibliography}
\end{document}